\documentclass[letterpaper]{IEEEtran}
\usepackage[T1]{fontenc}
\usepackage[latin9]{inputenc}
\usepackage[active]{srcltx}
\usepackage{xcolor}
\usepackage{float}
\usepackage{bm}
\usepackage{amsmath}
\usepackage{amsthm}
\usepackage{amssymb}
\usepackage{graphicx}

\makeatletter

\providecommand{\tabularnewline}{\\}
\floatstyle{ruled}
\newfloat{algorithm}{tbp}{loa}
\providecommand{\algorithmname}{Algorithm}
\floatname{algorithm}{\protect\algorithmname}

\theoremstyle{plain}
\newtheorem{thm}{\protect\theoremname}
\theoremstyle{remark}
\newtheorem{rem}[thm]{\protect\remarkname}
\def\R2#1{\textcolor{red}{#1}}

\usepackage{url}
\usepackage{psfrag}
\usepackage{flushend}
\usepackage[caption=false]{subfig}

\usepackage{algorithmic}
\usepackage{cite}

\@ifundefined{showcaptionsetup}{}{%
 \PassOptionsToPackage{caption=false}{subfig}}
\usepackage{subfig}
\makeatother

\providecommand{\remarkname}{Remark}
\providecommand{\theoremname}{Theorem}

\begin{document}

\title{{Location-Aided Pilot Contamination Avoidance for
Massive MIMO Systems}}

\author{L.~Srikar~Muppirisetty, Themistoklis~Charalambous,~\IEEEmembership{Member,~IEEE},
Johnny~Karout,~\IEEEmembership{Senior~Member,~IEEE}, G\'{a}bor~Fodor,~\IEEEmembership{Senior~Member,~IEEE},
and Henk~Wymeersch,~\IEEEmembership{Member,~IEEE} \thanks{L. Srikar Muppirisetty and Henk Wymeersch are with the Department
of Signals and Systems, Chalmers University of Technology, 41296 Gothenburg,
Sweden, e-mail: \{srikar.muppirisetty, henkw\}@chalmers.se.

Themistoklis Charalambous is with the Department of Electrical Engineering
and Automation,  School of Electrical Engineering, Aalto University,
Espoo, Finland email: themistoklis.charalambous@aalto.fi.

Johnny Karout and G\'{a}bor Fodor are with Ericsson Research, Kista,
16480 Stockholm, Sweden, email:\{johnny.karout, gabor.fodor\}@ericsson.com.

This research was supported, in part, by the European Research Council,
under Grant No. 258418 (COOPNET).{{} Part of this work
was presented in \cite{muppirisetty2015Location}. }}}
\maketitle
\begin{abstract}
{Pilot contamination, defined as the interference during the channel
estimation process due to reusing the same pilot sequences in neighboring cells,
can severely degrade the performance of massive
multiple-input multiple-output systems.}
{
In this paper, we propose
a location-based approach to mitigating the pilot contamination problem
for uplink multiple-input multiple-output systems. Our approach makes
use of the approximate locations of mobile devices to provide good
estimates of the channel statistics between the mobile devices and
their corresponding base stations. Specifically, we aim at avoiding
pilot contamination even when the number of base station antennas
is not very large, and when multiple users from different cells, or
even in the same cell, are assigned the same pilot sequence. First,
we characterize a desired angular region of the target user at the
serving base station based on the number of base station antennas
and the location of the target user, and make the observation that
in this region the interference is close to zero due to the spatial
separability. Second, based on this observation, we propose pilot
coordination methods for multi-user multi-cell scenarios to avoid
pilot contamination.
The numerical results indicate that the proposed
pilot contamination avoidance schemes enhance the quality of the
channel estimation} {and thereby improve}
{the per-cell sum rate offered
by target base stations.}
\end{abstract}


\section{Introduction}

The use of very large antenna arrays at the base station (BS) is considered
as a promising technology for 5G communications in order to cope with
the increasing demand of wireless services \cite{Boccardi2014Five}.{{}
Such massive multiple-input multiple-output (MIMO) systems provide
numerous advantages \cite{lim2013performance,larsson2014Massive,Rusek2013Sacling,Bjornson2014Optimizing,Marzetta2010Noncoop,Osseiran:16}:}
{including (i) increasing the spectral efficiency
by supporting a higher number of users per cell, (ii) improving energy
efficiency by radiating focused beams towards users, and (iii) averaging
out small scale fading resulting in the channel hardening effect}.
Furthermore, under the assumption of perfect channel estimation, massive
MIMO provides asymptotic orthogonality between vector channels of
the target and interfering users. However,  performance of these
systems is {degraded} by the pilot contamination effect, i.e., interference
during uplink channel estimation due to reusing of the same pilot
sequences.

Pilot sequences are a scarce resource due to the fact that the length
of pilot sequences (the number of symbols) is limited by the coherence
time and bandwidth of the wireless channel. As a result, the number
of separable users is limited by the number of the available orthogonal
pilot sequences \cite{Rusek2013Sacling,Bjornson2014Optimizing}. Therefore,
in multi-cell massive MIMO systems, the pilot sequences must be reused,
which leads to interference between identical pilot sequences from
users in either neighboring cells or even the same cell; this effect
is known as pilot contamination \cite{hoydis2013massive}.
Pilot contamination
is known to degrade the quality of channel state information at the
BS, which in turn degrades the performance in terms of the achieved
spectral efficiency, beamforming gains, and cell-edge user throughput.
{Since pilot contamination is an important phenomenon}
{that degrades the performance of massive MIMO systems, we provide a detailed
overview of existing works on this topic in the following subsection.}

\subsection{Related Works}

{Mitigation strategies for pilot contamination have
been well studied in the literature. A comprehensive survey on pilot
contamination in massive MIMO systems is provided in \cite{elijah2016comprehensive}.
The existing pilot decontamination methods for time division duplex
(TDD) MIMO systems are broadly grouped in to two categories: pilot-based
and subspace-based approaches.}

{In pilot-based approaches, each BS takes turns in
sending pilots in a non-overlapping fashion\cite{fernandes2013inter,marzetta,smith,quek,Zhang}.
In these works, the frame structure is modified such that the pilots
are transmitted in each cell in non-overlapping time slots \cite{marzetta,smith},
or pilots are transmitted in consecutive phases in which each BS keeps
silent in one phase and repeatedly transmits in other phases \cite{quek}.
Alternatively, a combination of downlink and uplink scheduled training can be used \cite{Zhang}.}

{Subspace-based approaches  exploit second-order
statistics and utilize covariance-aided channel estimation \cite{yin2013decontaminating,filippou2012decontaminating,Gesbert,Marzetta2010Noncoop,mueller2013blind,muller2013analysis,muller2014blind,cottatellucci2013analysis,Larsson,li2013spatial,you2015pilot}.
Second-order statistics of desired and interfering user channels are
exploited in \cite{Gesbert,Marzetta2010Noncoop,yin2013decontaminating,filippou2012decontaminating}.
The works in \cite{Gesbert,hoydis2013massive,Guiseppe,yin2013decontaminating,filippou2012decontaminating}
considered spatial correlated fading, while earlier works assumed
uncorrelated fading. The use of non-diagonal covariance matrices enables
the identification of spatially compatible users based on the spatial
correlation of the covariance matrices. A closed-form expression of
the non-asymptotic downlink rate exploiting the statistical second
order channel covariance matrices information is provided in \cite{li2013spatial}.
In \cite{mueller2013blind,muller2013analysis,cottatellucci2013analysis},
blind channel estimation with power and power-controlled hand-off
is studied with singular value decomposition of the received signal
matrix. The assumption that coherence time is larger than the number
of BS antennas is used in \cite{mueller2013blind,muller2013analysis,cottatellucci2013analysis}
and relaxed in \cite{muller2014blind}. Blind channel estimation using
eigenvalue-decomposition is described in \cite{Larsson}. }

{A game-theoretic approach to tackle pilot contamination
is studied in \cite{ahmadi2016game}. An iterative way of assigning
users to pilots is proposed such that in each iteration the minimum
signal-to-interference ratio is improved \cite{zhusmart2015}. In
\cite{Zhang:16}, users in each cell are grouped based on severity
of pilot contamination and a fractional pilot reuse scheme is employed.
Other approaches include, among others, greater-than-one pilot reuse
schemes \cite{Marzetta2010Noncoop,yang2013total,li2012non,saxenaMitigating2015,Li:13,you2015pilot,Zhang:16,Li:16}, and
location-aided pilot allocation schemes \cite{wang2015location,zhao2016location,akbar2016mitigating}.
In \cite{wang2015location}, a location-aided channel estimation method
is described, wherein to mitigate the inter-cell interference where
an FFT-based post-processing approach is employed after pilot-aided
channel estimation. In\cite{zhao2016location}, a location-aware novel
pilot assignment algorithm is proposed for heterogeneous networks.The
method ensures that users that are assigned to the same pilot sequence
have distinguishable angle-of-arrivals (AoAs) at the macro BSs, while
maintaining a large distances for the interfering users to the corresponding
small BSs. The work reported in \cite{akbar2016mitigating} proposed a location-aware
pilot assignment scheme for Rician channel models and exploited the
location-dependent line-of-sight (LOS) channel component during the
pilot assignment procedure. }

{
Recent insights about the impact of pilot contamination suggest that
the capacity of massive MIMO systems increases without bound as the number of antennas
tend to infinity, even in the presence
of pilot contamination, if proper precoding/combining are employed \cite{bjornson2017massive}.
As it is pointed out in that paper,
this result does not imply that the negative effects of pilot contamination disappear, since the resulting
estimation errors still degrade the performance.
As our numerical results indeed show,
the proposed pilot contamination avoidance scheme increases the sum rate of multicell systems.
}

\subsection{Contributions}

In this work, we build on \cite{Gesbert}, which focused on a greedy
pilot allocation for the asymptotic regime of infinite antennas at
the BS, where pilot contamination is fully eliminated. Greedy approaches
lead to performance degradation, as reported in \cite{saxenaMitigating2015},
while pilot contamination does not vanish for the practical finite
antenna regime. Here, we address both aspects and design explicitly
for the finite-antenna regime, while targeting overall system performance
by considering a joint design across multiple cells for all users
in the system. More specifically, the contributions of our paper are
as follows:
\begin{itemize}
\item {Similar to our previous work\cite{muppirisetty2015Location},
a location-aided approach is taken for pilot decontamination and we
relate mean and the standard deviation of the AoA to a user location,
rather than to a user's channel. }
\item We then propose an approach with which, in the presence of location
information of the users, we can quantify the effect of pilot contamination
for BSs with a finite number of antennas. This result helps us predict
how each user interferes with the rest of the users having identical
pilot sequences when  BSs are equipped with MIMO antennas, and the number
of antennas is not necessarily approaching infinity. This quantification
reveals that for a considerable number of antennas, there is a range
of angles for which the pilot contamination is very small.
\item Based on the above observation, we formulate pilot decontamination
as an integer quadratic programming problem that we are able to solve
for all the BSs as a joint optimization problem. In particular, we
propose multi-cell multi-user joint optimization problems such that
it takes into consideration during the pilot assignment the mutual
interference seen by the target users at their respective BSs.
\item {We propose a heuristic approach for assigning users
to BSs to decrease the computational complexity of the proposed joint
optimization schemes. The proposed heuristic algorithm exploits both
distance and AoA information of the users. }
\end{itemize}

\subsection{Outline}

The rest of the paper is structured as follows. In Section \ref{sec:System-Model},
we introduce the system model comprising the network model, the channel
model, the received pilot signal and MMSE estimator. Section \ref{sec:Pilot-Decontamination}
addresses the problem of pilot decontamination for BSs with (not necessarily
massive) MIMO antennas. In Section \ref{sec:Coordinated-Pilot-Assignment},
we present optimal user assignment and heuristic strategies for pilot
contamination avoidance under various configurations, building on
the theory developed in Section \ref{sec:Pilot-Decontamination}.
Section \ref{sec:Numerical-Results} demonstrates the performance
of the proposed methods and we compare them with other user selection
methods proposed in the literature. Finally, Section \ref{sec:Conclusions}
draws conclusions and discusses possible future directions.

\subsection*{Notation}

Throughout the paper, vectors are written in bold lower case letters
and matrices in bold upper case letters. For matrix $\mathbf{X}$,
matrices $\mathbf{X}^{\mathrm{T}}$ and $\mathbf{X}^{\mathrm{H}}$
denote its transpose and hermitian, respectively. The $i$-th entry
of vector $\mathbf{x}$ is denoted as $[\mathbf{x}]_{i}$. $\mathrm{vec}[\mathbf{X}]$
denotes stacking all the elements of $\mathbf{X}$ in a vector. $U[\mathcal{S}]$
denotes a uniform distribution over the intervals defined by the set
$\mathcal{S}$.{{} }A sequence of elements $\{a_{1},a_{2},\ldots\}$
is written in short as $\{a_{j}\}_{j}$. The positive operator is
denoted as $(x)^{+}=\max(0,x)$. The Kronecker product of two matrices
$\mathbf{X}_{1}$ and $\mathbf{X}_{2}$ is denoted as $\mathbf{X}_{1}\otimes\mathbf{X}_{2}$.
$\mathbf{I}_{M}$ denotes the identity matrix of size $M\times M$
and $\Vert\cdot\Vert_{2}$ denotes the Euclidean norm. The cardinality
of a set $\mathcal{A}$ is denoted by $|\mathcal{A}|$. The sets of
real and complex numbers are denoted by $\mathbb{R}$ and $\mathbb{C}$,
respectively; the $n$-dimensional Euclidean and complex spaces are
denoted by $\mathbb{R}^{n}$ and $\mathbb{C}^{n}$, respectively.

{Important angle symbols used in the paper are: $[\theta_{i}^{\mathrm{min}},\theta_{i}^{\mathrm{max}}]$ represents the AoA support of user $i$, where
$\theta_{i}^{\mathrm{min}}$ denotes the minimum AoA support angle and $\theta_{i}^{\mathrm{max}}$ denotes the maximum AoA support angle. The  mean AoA of user $i$ is denoted by $\theta_{i}^{\mu}$. The  AoA supports of the desired and interfering user after axis transformation are denoted by $I_{i}^{(i)}$ and $I_{j}^{(i)}$ respectively. The desired angular region of user $i$ is given by  $[\tilde{\psi}_{i}^{\mathrm{min}},\tilde{\psi}_{i}^{\mathrm{max}}]$.}

\section{System Model\label{sec:System-Model}}

\subsection{Network model \label{subsec:Network-model}}

We consider a two-dimensional scenario with cells, and each cell is
served by one BS equipped with $M$ antennas. We denote $\mathcal{C}$
as the set of all cells and where $\mathcal{K}_{j}$ the set of users
in the $j$-th cell, $j\in\mathcal{C}$. The set of neighboring cells
to the $j$-th cell is denoted by $\mathcal{C}_{j}^{\mathrm{sur}}$.
Users, equipped with a single antenna, are located uniformly within
the cells. The location of user $i\in\mathcal{K}_{j}$ is denoted
by $\mathbf{x}_{ij}\in\mathbb{R}^{2}$, while the location of the
BS in the $k$-th cell is written as $\mathbf{x}_{k}\in\mathbb{R}^{2}$.
We define $\mathcal{P}$ as the set of available orthogonal pilots
for allocation to users.

\subsection{Channel Model\label{subsec:Channel-Model}}

The uplink channel of user $i$ from cell $j$ to BS $k$ is denoted
by $\mathbf{h}_{ijk}\in\mathbb{C}^{M}$. We note that the channel
depends only on user $i$ and BS $k$, but the use of the additional
index $j$ allows us to distinguish in which cell the users belong
to. {We consider the channel as the superposition
of a large number of paths \cite{Gesbert,Guiseppe,tsaitheimpact2002}:}
\begin{equation}
\mathbf{h}_{ijk}=\sqrt{\frac{\beta_{ijk}}{B}}\sum_{b=1}^{B}\mathbf{a}(\theta_{ijk}^{(b)})\,\alpha_{ijk}^{(b)},\label{eq:channel_eq}
\end{equation}
where $\beta_{ijk}=\alpha\Vert\mathbf{x}_{ij}-\mathbf{x}_{k}\Vert_{2}^{-\eta}$,
for path-loss exponent $\eta$ and a known\footnote{The constant $\alpha$ depends on cell-edge signal-to-noise ratio
(SNR), as specified in the numerical results. } constant $\alpha$, $\mathbf{a}(\theta_{ijk}^{(b)})\in\mathbb{C}^{M}$
is the antenna steering vector corresponding to AoA $\theta_{ijk}^{(b)}\in[0,2\pi)$,
$b$ is the path index, and $\alpha_{ijk}^{(b)}$ is the random phase
of the $b$-th path. We restrict ourselves to uniform linear arrays
with $[\mathbf{a}(\theta_{ijk})]_{m}=\exp(-j2\pi mD\cos(\theta_{ijk})/\lambda)$,
for the antenna spacing $D$ and the signal wavelength $\lambda$.{{}
Under the assumption of $B\to+\infty$ and i.i.d.~AoAs with common
probability density $p(\theta_{ijk})$, application of the law of
large numbers (center limit theorem) gives rise to $\mathbf{h}_{ijk}$ having a zero-mean
Gaussian distribution with covariance matrix
\begin{align}
\mathbf{R}_{ijk} & =\mathbb{E}\Bigl[\mathbf{h}_{ijk}\,\bigl(\mathbf{h}_{ijk})^{\mathrm{H}}\Bigr]\\
 & =\beta_{ijk}\int_{0}^{2\pi}p(\theta_{ijk})\mathbf{a}(\theta_{ijk})\mathbf{a}^{\mathrm{H}}(\theta_{ijk})\,\mathrm{d}\theta_{ijk}.
\end{align}
We further assume that $p(\theta_{ijk})$ corresponds to a uniform
distribution with support $[\theta_{ijk}^{\mathrm{min}},\theta_{ijk}^{\mathrm{max}}]$,
for some fixed $\theta_{ijk}^{\mathrm{min}},\theta_{ijk}^{\mathrm{max}}\in[0,2\pi]$,
$\theta_{ijk}^{\mathrm{min}}<\theta_{ijk}^{\mathrm{max}}$ . Finally,
we assume that a map exists in the BS, associating the user's position
to the support of the AoA distribution as well as the average received
power (i.e., $\beta_{ijk}$ in the form of a radio map).}
\begin{rem}
{In some scenarios, especially when the BS is elevated
and seldom obstructed, propagation can be dominated by scatterers
in the vicinity of the users, giving rise a limited AoA spread \cite{Guiseppe,Shiu:00,Abdi:02,Oestges:03,Oestges:04,Zhang:07},
as assumed in this work. We note that the assumption of a uniform
AoA distribution can be generated with the widely used ring model
\cite{tsaitheimpact2002,Gesbert,Gao2015Spatial,Guiseppe,Fang2017Low}.
Such a ring model will be used for illustration and  to exemplify
the approach, } {but is not replied upon in the development of the method.}
\end{rem}

\subsection{Received Pilot Signal and MMSE Channel Estimator\label{subsec:Received-Signal}}

The target user $i$ in cell $k$ sends uplink transmission to BS
$k$. Users from different cells have been assigned the same pilot
sequence $\mathbf{s}$ of length $\tau$. For notational convenience,
we will assume that all the users indexed with $i$ are also assigned
the same pilot sequence $\mathbf{s}$. Later, in Section \ref{sec:Coordinated-Pilot-Assignment},
we will present various ways to assign users across cells to a given
pilot sequence. The received $M\times\tau$ pilot signal observed
at BS $k$ is written as
\begin{equation}
\mathbf{Y}_{k}=\mathbf{h}_{ikk}\,\mathbf{s}^{\mathrm{T}}+\sum_{j\in\mathcal{C}}\mathbf{h}_{ijk}\,\mathbf{s}^{\mathrm{T}}+\mathbf{N},\label{eq:received_signal}
\end{equation}
where $\mathbf{N}\in\mathbb{C}^{M\times\tau}$ is spatially and temporally
additive white Gaussian noise (AWGN) with element-wise variance $\sigma^{2}$
. In (\ref{eq:received_signal}), $\mathbf{h}_{ikk}$ is the desired
signal channel in the cell $k$ and $\mathbf{h}_{ijk}$ are the channels
of interfering users.

The MMSE estimate of the desired channel $\mathbf{h}_{ikk}$ by BS
$k$ is given by \cite[Eq. (18)]{Gesbert}
\begin{equation}
\hat{\mathbf{h}}_{ikk}=\mathbf{R}_{ikk}\Bigl(\sigma^{2}\mathbf{I}_{M}+\tau(\mathbf{R}_{ikk}+\sum_{j\in\mathcal{C}}\mathbf{R}_{ijk})\Bigr)^{-1}\bar{\mathbf{S}}^{\mathrm{H}}\mathrm{vec}[\mathbf{Y}_{k}],\label{eq:MMSEinterference}
\end{equation}
where $\bar{\mathbf{S}}=\mathbf{s}\otimes\mathbf{I}_{M}$ and $\mathbf{R}_{ijk}\in\mathbb{C}^{M\times M}$
is the covariance matrix of $\mathbf{h}_{ijk}$.

\section{Pilot Decontamination\label{sec:Pilot-Decontamination} }

In this section, we discuss pilot decontamination methods under massive
and finite antenna array setting. For legibility, the extra subscripts
$k$ are dropped. We consider the scenario of a given target user,
say user $i$, {with channel $\mathbf{h}_{i}$ that has arbitrary AoA distribution,
but has a support $[\theta_{i}^{\mathrm{min}},\theta_{i}^{\mathrm{max}}]$}. Our objective
is to find an interfering user, say user $j,$ in the surrounding
cells and assign it the same pilot sequence as user $i$ in such a
way as to minimize interference during channel estimation for user
$i$. These users have AoAs in the ranges $\{[\theta_{j}^{\mathrm{min}},\theta_{j}^{\mathrm{max}}]\}_{j}$
with respect to (w.r.t.) target BS for the corresponding channels
$\{\mathbf{h}_{j}\}_{j}$.{{} It will turn out to be
convenient to make the target BS as origin and mean AoA $\theta_{i}^{\mu}$
of user $i$ w.r.t. to BS be the new zero-degree axis. All other users
are transformed according to the new coordinate system. In particular,
we apply the axis transformation so that $\theta_{i}^{\mu}$ is the
new zero-degrees axis.} The corresponding modified AoA supports of
the desired and interfering user after axis transformation are $I_{i}^{(i)}$
and $I_{j}^{(i)}$ respectively. Furthermore, we denote $\theta_{i}^{(i)}\in I_{i}^{(i)}$
and $\theta_{j}^{(i)}\in I_{j}^{(i)}.$ The subscript denotes the
user index and the superscript indicates with respect to which user
the axis transformation has been applied.
\begin{figure*}
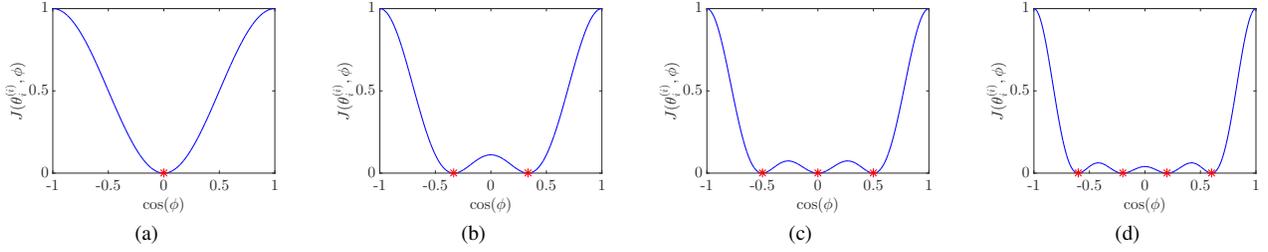

\begin{centering}
\subfloat[]{\begin{centering}
\includegraphics[width=0.2\textwidth,height=0.2\textheight,keepaspectratio]{J_phi_M_2}
\par\end{centering}
}\quad\subfloat[]{\begin{centering}
\includegraphics[width=0.2\textwidth,height=0.2\textheight,keepaspectratio]{J_phi_M_3}
\par\end{centering}
}\quad\subfloat[]{\begin{centering}
\includegraphics[width=0.2\textwidth,height=0.2\textheight,keepaspectratio]{J_phi_M_4}
\par\end{centering}
}\quad\subfloat[]{\begin{centering}
\includegraphics[width=0.2\textwidth,height=0.2\textheight,keepaspectratio]{J_phi_M_5}
\par\end{centering}
}
\par\end{centering}
\caption{\label{fig:J_theta_various_M} The behavior $J(\theta_{i}^{(i)}\phi)$
with $\cos(\phi)$, for $\theta_{i}^{(i)}=0$, $\beta_{i}=1$, and
for various antenna array lengths: (a) $M=2$, (b) $M=3$, (c) $M=4$,
(d) $M=5$. The zeros $\{\cos(\phi_{r}^{*})\}_{r}$\} are depicted
with red asterisks ({*}).{{} }}
\end{figure*}

\subsection{Massive MIMO}
{Note that with the term ``massive MIMO'', we also refer to
finite-dimensional antenna systems, similarly to most of the literature \cite{Huh:12,Rusek2013Sacling,Ngo:13}.}
For a massive antenna array setting, it has been shown in \cite[Theorem 1]{Gesbert}
that when the intervals $\{I_{j}^{(i)}\}_{j}$ are strictly non-overlapping
with $I_{i}^{(i)}$, i.e., $I_{j}^{(i)}\cap I_{i}^{(i)}=\emptyset,\forall j\neq i$, then {the channel estimate $\hat{{\bf h}}_{i}$ tends to the interference free channel estimate $\hat{{\bf h}}_{i}^{\textrm{no-int}}$,}
i.e., the channel estimate
of the desired channel in the presence of no interfering pilot signals
from other cells, obtained by setting the interference terms to zero
in (\ref{eq:MMSEinterference}), leading to $\hat{{\bf h}}_{i}^{\textrm{no-int}}=\mathbf{R}_{i}(\sigma^{2}\mathbf{I}_{M}+\tau\mathbf{R}_{i})^{-1}\bar{\mathbf{S}}^{\mathrm{H}}\mathbf{y},$
where $\mathbf{y}=\bar{\mathbf{S}}\mathbf{h}_{i}+\mathrm{vec}[\mathbf{N}]$
is the received $\tau\times1$ pilot signal vector at the BS under
no interference from the other cell users.
{In fact,
a sum-rate performance close to that of the interference-free
channel estimation scenario is obtained for finite numbers
of antennas and users. As the number of
antennas goes to infinity, the pilot contamination is completely eliminated
and the channel estimate of the target user reaches the interference-free
scenario.}

\subsection{Finite MIMO\label{subsec:Finite-MIMO}}

{Inspired by the approach in \cite{Gesbert},
we carry out the analysis that is applicable for moderate and large number of antennas $M$}.
Our objective is to have a cost measure to the channel estimation
quality of user $i$ when user $j$ is assigned the same pilot sequence.

\subsubsection{Condition for limited interference}

Let us consider the desired user AoA after axis transformation is
bounded by $p(\theta_{i}^{(i)})$.{{} }We aim to determine
for which AoAs a user $j$ causes only a small degradation to the
channel estimation of user $i$. In Appendix \ref{sec:Channel-estimation-finiteMIMO},
we show the interference from user $j$ with normalized steering vector
$\mathbf{a}(\phi)/\sqrt{M}$ is small when $\mathbf{a}^{\mathrm{H}}(\phi)\mathbf{R}_{i}\mathbf{a}(\phi)/M$
is small. Hence, users with steering vectors $\mathbf{a}(\phi)$ for
which
\begin{align}
\frac{\mathbf{a}(\phi)^{\mathrm{H}}}{\sqrt{M}}\mathbf{R}_{i}\frac{\mathbf{a}(\phi)}{\sqrt{M}} & =\frac{1}{M}\mathbb{E}\Bigl[\left|{\sqrt{{\beta_{i}}}}\mathbf{a}(\phi)^{\mathrm{H}}\mathbf{a}(\theta_{i}^{(i)})\right|^{2}\Bigr]\\
 & =\frac{1}{M}\int J^{2}(\theta_{i}^{(i)},\phi)\,p(\theta_{i}^{(i)})\,\mathrm{d}\theta_{i}^{(i)}\label{eq:metricforlowinterference}
\end{align}
is small will lead to limited degradation during channel estimation
of user $i$. We have introduced
%
{
\begin{align}
 & J(\theta_{i}^{(i)},\phi)=\label{eq:JfunctionDef}\\
 & \sqrt{\beta_{i}}\left|\sum_{m=1}^{M}\exp(2\pi j(m-1)\frac{D}{\lambda}(\cos(\phi)-\cos(\theta_{i}^{(i)})))\right|\nonumber \\
 & =
\begin{cases}
\sqrt{\beta_{i}}\frac{\Bigl|1-\exp\Bigl(2\pi jM\frac{D}{\lambda}(\cos(\phi)-\cos(\theta_{i}^{(i)}))\Bigr)\Bigr|}{\Bigl|1-\exp\Bigl(2\pi j\frac{D}{\lambda}(\cos(\phi)-\cos(\theta_{i}^{(i)}))\Bigr)\Bigr|}, & \cos(\phi)\neq \cos(\theta_{i}^{(i)}) \\[0.3cm]
\sqrt{\beta_{i}}M, & \text{otherwise.}
\end{cases}
\label{eq:geometricinterpretation}
\end{align}}

Note that when $M\rightarrow\infty$ and $\phi\notin I_{i}^{(i)}$,
then $J^{2}(\theta_{i}^{(i)},\phi)/M\to0$, as shown in \cite{Gesbert},
so that (\ref{eq:metricforlowinterference}) also vanishes. In contrast,
for finite $M$, the shape of $J^{2}(\theta_{i}^{(i)},\phi)/M$ must
be accounted for. {We note that
\begin{align*}
 & \frac{\mathbf{a}(\phi)^{\mathrm{H}}}{\sqrt{M}}\mathbf{R}_{i}\frac{\mathbf{a}(\phi)}{\sqrt{M}}\\
 & \le\frac{1}{M}\int\max_{\theta}\left[J^{2}(\theta,\phi)\right]p(\theta_{i}^{(i)})\,\mathrm{d}\theta_{i}^{(i)}\\
 & =\frac{\max_{\theta}J^{2}(\theta,\phi)}{M},
\end{align*}
so that users with AoA $\phi$ are preferred when $\max_{\theta_{i}^{(i)}}J^{2}(\theta_{i}^{(i)},\phi)$
is small, or equivalently, when $\max_{\theta_{i}^{(i)}}J(\theta_{i}^{(i)},\phi)$,
is small. The region of $\phi$ for which
\[
J_{i}^{\mathrm{DAR}}(\phi)=\max_{\theta_{i}^{(i)}}J(\theta_{i}^{(i)},\phi)
\]
 is small is termed the desired angular region (DAR). To define the
DAR, we must understand the behavior of $J(\theta_{i}^{(i)},\phi)$
as a function of $M$.}\textcolor{red}{}

\subsubsection{Characterization of $J(\theta_{i}^{(i)},\phi)$}

We characterize $J(\theta_{i}^{(i)},\phi)$ through its minima and
maxima. It can be easily deduced that the minimum of $J(\theta_{i}^{(i)},\phi)$
is $0$ and is attained when the numerator becomes zero, i.e.,
\begin{align}
\exp\Bigl(2\pi jM\frac{D}{\lambda}(\cos(\phi)-\cos(\theta_{i}^{(i)}))\Bigr)=1,\label{eq:phi_star1}
\end{align}
which is equivalent to
\begin{align}
\cos(\phi)=\cos(\theta_{i}^{(i)})+\frac{z\lambda}{MD},\label{eq:cosf}
\end{align}
where $z\in\mathbb{Z}$, such that $\cos(\phi)\in[-1,1]$ and $\phi\notin I_{i}^{(i)}$.
Fig.~\ref{fig:J_theta_various_M} depicts the behavior of $J(\theta_{i}^{(i)},\phi)$
when $\theta_{i}^{(i)}=0$ and for various values of $M$, where $J(\theta_{i}^{(i)},\phi)$
is computed numerically for all values of $\phi$ using (\ref{eq:geometricinterpretation}).
It can be observed that the number of zeros of the function $J(\theta_{i}^{(i)},\phi)$
depends on the number of antennas and it is equal to $M-1$.
\begin{figure}[h]
\includegraphics[width=0.999\columnwidth]{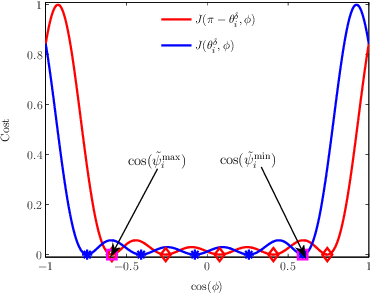} \caption{\label{fig:example1}The behavior of the cost functions $J(\theta_{i}^{\delta},\phi)$
and $J(\pi-\theta_{i}^{\delta},\phi)$ for the values of $\cos(\phi)=[-1,1]$,
$M=6$, $\theta_{i}^{\delta}=\frac{\pi}{8}$, and $\beta_{i}=1$.
Red diamonds and blue asterisks represent the zeros of the functions
$J(\theta_{i}^{\delta},\phi)$, $J(\pi-\theta_{i}^{\delta},\phi)$,
respectively. Purple squares denote $\tilde{\psi}_{i}^{\min}$ and
$\tilde{\psi_{i}}^{\max}$. }
\end{figure}
\begin{figure}
\includegraphics[width=1\columnwidth]{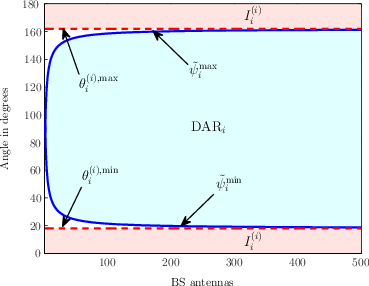}\caption{\label{fig:Relation_btw_psi_theta}The relation between $\tilde{\psi_{i}}^{\mathrm{min}}$
to $\theta_{i}^{(i),\min}$ and $\tilde{\psi_{i}}^{\mathrm{max}}$
to $\theta_{i}^{(i),\max}${{} }with increasing BS antennas
for $\theta_{i}^{\delta}=\frac{\pi}{10}$. Asymptotically, when $M\rightarrow\infty$,
$\tilde{\psi_{i}}^{\min}$ converges to $\theta_{i}^{(i),\min}$ and
$\tilde{\psi_{i}}^{\max}$ to $\theta_{i}^{(i),\max}$. The shaded
regions are: $\textrm{DAR}_{i}$ (blue) and $I_{i}^{(i)}$ (red).}
\end{figure}

To ensure limited impact of user $j$ on user $i$, $J(\theta_{i}^{(i)},\phi)$
should be small for all $\phi\in I_{j}^{(i)}$ and all $\theta_{i}^{(i)}\in I_{i}^{(i)}$.
Hence, we consider $J(\theta_{i}^{(i)},\phi)$ for $\theta_{i}^{(i)}$
at the boundaries of $I_{i}^{(i)},$ i.e., $\theta_{i}^{(i)}=\theta_{i}^{\delta}$
and $\theta_{i}^{(i)}=\pi-\theta_{i}^{\delta}$, as shown in Fig.~\ref{fig:example1}.
We observe that both $J(\theta_{i}^{\delta},\phi)$ and $J(\pi-\theta_{i}^{\delta},\phi)$
are small for $\cos(\phi)\in[\cos(\tilde{\psi_{i}}^{\min}),\cos(\tilde{\psi_{i}}^{\max})]$,
where the values of $\tilde{\psi_{i}}^{\min}$ and $\tilde{\psi_{i}}^{\max}$
are detailed in Appendix \ref{sec:Selection-of-} and visualized in
Fig.~\ref{fig:example1}.

Therefore, when the AoA support $I_{j}^{(i)}$ of the interfering
user $j$ lies within the desired angular region ($\textrm{DAR}_{i}$)
$[\tilde{\psi}_{i}^{\mathrm{min}},\tilde{\psi}_{i}^{\mathrm{max}}]$
of the target user $i$ then the interference is limited. This property
is exploited to devise various coordinated pilot assignment schemes
in Section \ref{sec:Coordinated-Pilot-Assignment}. As it is observed
in Fig. \ref{fig:J_theta_various_M}, the range of the $\textrm{DAR}_{i}$
increases with BS antennas. The relation between $\tilde{\psi_{i}}^{\mathrm{min}}$,
$\theta_{i}^{(i),\min}$ and $\tilde{\psi_{i}}^{\mathrm{max}}$, $\theta_{i}^{(i),\max}$
with $M$ is depicted in Fig. \ref{fig:Relation_btw_psi_theta}. Asymptotically,
when $M\rightarrow\infty$, we note that $\lim_{M\rightarrow\infty}\tilde{\psi_{i}}^{\min}=\theta_{i}^{(i),\min}=\theta_{i}^{\delta}$
and $\lim_{M\rightarrow\infty}\tilde{\psi_{i}}^{\max}=\theta_{i}^{(i),\max}=\pi-\theta_{i}^{\delta}$.
In \cite{Gesbert}, only for the case $M\rightarrow\infty$, the AoA
condition that needs to be satisfied between interfering and target
users is provided, while Fig. \ref{fig:Relation_btw_psi_theta} complements
this for finite $M$.
\begin{rem}
{For AoA distributions that are not bounded, the distribution
can be bounded by truncating the support. For AoA distributions that
can be described by union of multiple uniform distributions, the approach
can be generalized by introducing multiple disjoint DARs.}
\end{rem}

\subsubsection{Approximation of cost function}

The function $J_{i}^{\mathrm{DAR}}(\phi)$ is shown in Fig.~\ref{fig:Approximation}
and can be approximated by a piecewise linear function $J_{i}^{\mathrm{Apprx}}(\phi)$
(See blue dotted line in Fig. \ref{fig:Approximation}):
\begin{figure}
\includegraphics[width=1\columnwidth]{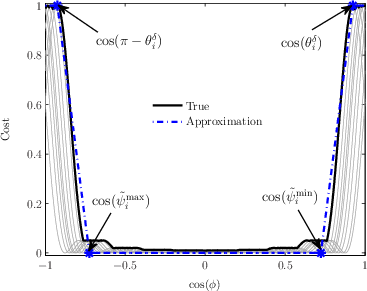}\caption{\label{fig:Approximation}The comparison between true cost function
$J_{i}^{\mathrm{DAR}}(\phi)$ and the approximate cost function $J_{i}^{\mathrm{Apprx}}(\phi)$
for $M=10$, $\theta_{i}^{\delta}=\frac{\pi}{8}$, and $\beta_{i}=1$.
The approximate cost function $J_{i}^{\mathrm{Apprx}}(\phi)$ is a
piece wise linear function connecting the following points $(-1,1),(\cos(\pi-\theta_{i}^{\delta}),1),(\cos(\tilde{\psi}_{i}^{\mathrm{max}}),0),(\cos(\tilde{\psi}_{i}^{\mathrm{min}}),0),(\cos(\theta_{i}^{\delta}),1),(1,1)$.
The thin grey lines represent the cost functions $J(\theta_{i}^{(i)},\phi),\forall\theta_{i}^{(i)}\in I_{i}^{(i)}$.}
\end{figure}
\begin{align}
 & J_{i}^{\mathrm{Apprx}}(\phi)=\sqrt{{\beta_{i}}}\times\\
 & \begin{cases}
1, & \cos(\phi)\leq\cos(\pi-\theta_{i}^{\delta}),\\
1-\frac{\cos(\phi)+\cos(\theta_{i}^{\delta})}{\cos(\tilde{\psi}_{i}^{\mathrm{max}})+\cos(\theta_{i}^{\delta})}, & -\cos(\theta_{i}^{\delta})\leq\cos(\phi)\leq\cos(\tilde{\psi}_{i}^{\mathrm{max}}),\\
\frac{\cos(\phi)-\cos(\tilde{\psi}_{i}^{\mathrm{min}})}{\cos(\theta_{i}^{\delta})-\cos(\tilde{\psi}_{i}^{\mathrm{min}})}, & \cos(\tilde{\psi}_{i}^{\mathrm{min}})\leq\cos(\phi)\leq\cos(\theta_{i}^{\delta}),\\
1, & \cos(\phi)\geq\cos(\theta_{i}^{\delta}),\\
0, & \mathrm{elsewhere}.
\end{cases}\nonumber
\end{align}
Based on the $J^{\mathrm{Apprx}}(\phi)$, we define $J_{ij}$ as the
interference cost to the BS of user $i$ experiences from another
user $j$, which basically\textbf{ }assigns zero cost when $I_{j}^{(i)}$
lies within the $\textrm{DAR}_{i}$ of user\textbf{ }$i$. Outside
the $\textrm{DAR}_{i}$, the cost grows linearly and saturates to
$\sqrt{{\beta_{i}}}$ at $\theta_{i}^{(i),\min}$ and
$\theta_{i}^{(i),\max}.$

\subsubsection{Cost of pilot assignment}

Based on the notion of the DAR and the function $J_{i}^{\mathrm{Apprx}}(\phi)$,
we can finally determine a cost to user $i$ when user $j$ is assigned
the same pilot. In particular, for the interfering user $j$ with
$I_{j}^{(i)}=[\theta_{j}^{(i),\min},\theta_{j}^{(i),\max}],\theta_{j}^{(i),\min}<\theta_{j}^{(i),\max}$,
we can introduce
\[
J_{ij}=J_{i}^{\mathrm{Apprx}}(\theta_{j}^{(i),\min})+J_{i}^{\mathrm{Apprx}}(\theta_{j}^{(i),\max}).
\]
The interference cost $J_{ij}$ is used in devising coordinated pilot
assignment schemes described in Section \ref{sec:Coordinated-Pilot-Assignment}.

\section{Coordinated Pilot Assignment Schemes\label{sec:Coordinated-Pilot-Assignment}}

In this section, we describe four different user assignment strategies
for pilot decontamination under various configurations based on the
theory developed in Section \ref{sec:Pilot-Decontamination}.

\subsection{Multi-User Multi-Cell Optimization\label{subsec:Multi-Cell} }

For this scenario, the goal is to reuse the pilots among the users
in the best possible way such that each user has been allocated a
pilot. Let us collect the users from all the cells in a set $\mathcal{N}=\cup_{j\in\mathcal{C}}\,\mathcal{K}_{j}$,
where $\mathcal{C}$ denotes the set of all cells and $\mathcal{K}_{j}$
the set of users in the $j$-th cell, $j\in\mathcal{C}$. Recall that
$\mathcal{P}$ is the set of all available orthogonal pilot sequences.
Let us introduce the variable $y_{ip}\in\{0,1\}$, with $y_{ip}=1$,
if $\ensuremath{i}$-th user is activated on $p$-th pilot and 0 otherwise.
The one-shot joint optimization for user assignment for multi-user
and multi-cell scenario can be written as a binary integer program
(BIP):\begin{subequations}\label{multi_user_multi_cell}
\begin{align}
\textrm{minimize\,\,\,} & \sum_{p\in\mathcal{P}}\sum_{i\in\mathcal{N}}\sum_{j\neq i}U_{ij}y_{ip}y_{jp}\label{eq:multi_user_obj-1}\\
 & \sum_{p\in\mathcal{P}}y_{ip}\geq1,\,\forall i\in\mathcal{N}\label{eq:multi_user_const-1}\\
 & y_{ip}\in\{0,1\},\,\forall i\in\mathcal{N},\,p\in\mathcal{P},\label{eq:multi_user_const-2}
\end{align}
\end{subequations}where
\begin{align}
U_{ij} & =\begin{cases}
J_{ij}, & i\neq j,\\
0, & i=j,
\end{cases}\label{eq:Uij}
\end{align}

We note the following: (\ref{eq:multi_user_obj-1}) gives preference
to users who lie within the desired angular region of the target user;
(\ref{eq:multi_user_const-1}) states each user must be active at
least on one pilot; and (\ref{eq:multi_user_const-2}) imposes the
binary integer requirements on the optimization variables. For each
pilot, the optimization (\ref{multi_user_multi_cell}), looks for
users in each cell with users in every cell and then chooses a set
of users such that when assigned the same pilot to them, they will
have minimum possible interference at their respective BSs.

The user assignment is performed based on the location for a target
user, say $i$, accounting for the following cases:
\begin{description}
\item [{$\textbf{(C1)}$}] The support of interfering signals AoA $I_{j}^{(i)},\forall j$
lies \emph{completely inside} the $\textrm{DAR}_{i}$ of the target
user;
\item [{$\textbf{(C2)}$}] The support of interfering signal AoA $I_{j}^{(i)},\forall j$
lies \emph{partially inside} the $\textrm{DAR}_{i}$ of the target
user; and
\item [{$\textbf{(C3)}$}] The support of interfering signal AoA $I_{j}^{(i)},\forall j$
lies \emph{completely outside} with the $\textrm{DAR}_{i}$ of the
target user.
\end{description}
The above optimization problem is always feasible. The problem (\ref{multi_user_multi_cell})
gives preference to users that satisfy the case \textbf{(C1)}, in
which case the objective is zero. It might be possible that the user
locations are such that \textbf{(C1)} cannot be satisfied. This is
tackled in (\ref{multi_user_multi_cell}), as it implicitly considers
the cases \textbf{(C2)} and \textbf{(C3)} in the formulation. For
example, when $I_{j}^{(i)}$ does not lie within $\textrm{DAR}_{i},$
then the objective function becomes positive. Therefore, to minimize
the objective, the interfering users are selected in such a way that
the maximal overlap with the desired support of the target user is
obtained.

The optimization (\ref{multi_user_multi_cell}) can be written as
an integer quadratic constraint optimization problem (IQCP) as\begin{subequations}\label{multi_user_multi_cell_Quad}
\begin{align}
\textrm{minimize}\,\,\, & \mathbf{y}^{\mathrm{T}}\mathbf{Q}\,\mathbf{y}\\
 & \sum_{p\in\mathcal{P}}y_{ip}\geq1,\forall i\in\mathcal{N}\label{eq:constraint_each_user_atleast_one_pilot}\\
 & y_{ip}\in\{0,1\},\,\forall i\in\mathcal{N},\,p\in\mathcal{P},
\end{align}
\end{subequations}where $\mathbf{y}=[y_{11},\ldots,y_{|\mathcal{N}|1},\ldots,y_{1|\mathcal{P}|},\ldots,y_{|\mathcal{N}||\mathcal{P}|}]$,
$\mathbf{Q}=\mathbf{I}_{|\mathcal{P}|}\otimes\mathbf{U}$ is a $|\mathcal{N}||\mathcal{P}|\times|\mathcal{N}||\mathcal{P}|$
block diagonal matrix where $\mathbf{U}$ is the $|\mathcal{N}|\times|\mathcal{N}|$
utility matrix with entries
\begin{equation}
\mathbf{U}=\left[\begin{array}{cccc}
U_{11} & U_{12} & \cdots & U_{1|\mathcal{N}|}\\
U_{21} & U_{22} & \cdots & U_{2|\mathcal{N}|}\\
\vdots & \vdots & \ddots & \vdots\\
U_{|\mathcal{N}|1} & U_{|\mathcal{N}|2} & \cdots & U_{|\mathcal{N}||\mathcal{N}|}
\end{array}\right],
\end{equation}
where $U_{ij}$ is given in \eqref{eq:Uij}.
\begin{rem}
The optimization problem \eqref{multi_user_multi_cell} is formulated
as an IQCP and it can be easily shown to be NP-hard. {More
specifically, quadratic problems (QPs) are known to be NP (see for
example, \cite{VAVASIS:1990}) and only the convex QP admits a polynomial
time solution \cite{KOZLOV1980223}. The standard QP problem where
the constraints are linear, is NP-hard if matrix $\mathbf{Q}$ is
indefinite \cite{Sahni:1974}. In our case, the structure of matrix
$\mathbf{Q}$ is non-symmetric, since in general $J_{ij}$ is not
necessarily equal to $J_{ji}$, and hence it is indefinite. Also,
in our case, the Boolean constraints are nonconvex, since the IQCP
can be written as a quadratic constrained quadratic programming (QCQP)
in which $y_{ip}(1-y_{ip})=0$.}

It should be also emphasized that the above optimization problem is
dependent on the number of antennas at each BS via $U_{ij}$ (see
\eqref{eq:Uij}).{{} Furthermore, the optimization relies
on the fact that BSs acquire} {location information
of users, which in turn incurs an overhead. Our assumption is that
inter-BS communication using, for example the X2 interface in Long
Term Evolution (LTE) systems, on the time scale of 100 ms would make
such overhead for the inter-BS communication tolerable in a real system.}

The formulation of the optimization problem \eqref{multi_user_multi_cell_Quad}
is very general and encompasses a multitude of possible scenarios
even when the users for assignment belong to the same cell as that
of the target user. In what follows, we show some variants of interest
of \eqref{multi_user_multi_cell_Quad} in the subsequent sections.
\end{rem}

\subsection{Multi-User Multi-Cell Optimization with QoS Guarantees\label{subsec:Multi-User-Multi-Cell-Optimizati}}

For the sake of establishing QoS guarantees, we may want to exclude
the possibility of assigning users from the same cell as that of the
target user to the same pilot. This is achieved by changing \eqref{eq:constraint_each_user_atleast_one_pilot}
to a constraint such that only one user from each cell is assigned
per pilot; this is enforced in the optimization problem via \eqref{eq:QoS_one_per_cell}.
Note that if no user exists in a cell for a certain pilot, then for
constraint \eqref{eq:QoS_one_per_cell} to be valid, without loss
of generality, the constraint for that cell at a certain pilot is
removed. Therefore, the IQCP formulation for multi-user and multi-cell
optimization with QoS guarantees is written as \begin{subequations}\label{multi_user_multi_cell_qos}
\begin{align}
\textrm{minimize}\,\,\, & \mathbf{y}^{\mathrm{T}}\mathbf{Q}\,\mathbf{y}\\
 & \sum_{i\in\mathcal{K}_{j}}y_{ip}=1,\forall j\in\mathcal{C},\,p\in\mathcal{P}\label{eq:QoS_one_per_cell}\\
 & y_{ip}\in\{0,1\},\,\forall i\in\mathcal{N},\,p\in\mathcal{P}.
\end{align}

\end{subequations}

\subsection{Multi-User Single-Cell Optimization}

In this scenario, our objective is to find for each user in a target
cell one user from the neighboring cells and assign them the same
pilot sequence. For this scenario, the mutual interference of the
users at the respective BSs is ignored and only the interference of
the users observed at the target cell are required to be satisfied.
This scenario finds applications to cases where priority is required
to be given to a certain cell (for example, in case there is a special
event that requires wireless communications to be robust) and in dense
urban areas where it is computationally very expensive to include
all the cells in the network and inevitably to run such a large-scale
optimization in real time.

The set of users for this scenario are the users from the target cell
and its neighboring cells. Let the set of users given by $\mathcal{M}=\mathcal{K}_{q}\cup(\cup_{i\in\mathcal{C}_{q}^{\mathrm{sur}}}\mathcal{K}_{i}$),
where $\mathcal{C}_{q}^{\mathrm{sur}}$ is the set of cells surrounding
cell $q.$ The modified optimization is then written as\begin{subequations}\label{multi_user_single_cell}
\begin{align}
\textrm{minimize}\,\,\, & \mathbf{y}^{\mathrm{T}}\bar{\mathbf{Q}}\,\mathbf{y}\\
 & \sum_{i\in\mathcal{K}_{j}}y_{ip}=1,\forall j\in(q\cup\mathcal{C}_{q}^{\mathrm{sur}}),\,p\in\mathcal{P}\\
 & y_{ip}\in\{0,1\},\,\forall i\in\mathcal{M},\,p\in\mathcal{P},
\end{align}
\end{subequations}where $\bar{\mathbf{Q}}=\mathbf{I}_{|\mathcal{P}|}\otimes\bar{\mathbf{U}}$
is an $|\mathcal{M}||\mathcal{P}|\times|\mathcal{M}||\mathcal{P}|$
block diagonal matrix where $\mathbf{\mathbf{\bar{U}}}$ is the $|\mathcal{M}|\times|\mathcal{M}|$
dimension utility matrix with entries
\begin{equation}
\bar{U}_{ij}=\begin{cases}
U_{ij} & ,\:\textrm{if }i\in\mathcal{K}_{q},\\
0 & ,\:\text{otherwise.}
\end{cases}
\end{equation}
\textcolor{cyan}{}

\subsection{{Smart Pilot Assignment\cite{zhusmart2015}\label{subsec:Smart-Pilot-Algorithm}}}

{In this subsection, a modified smart pilot assignment
of \cite{zhusmart2015} for user assignment is described. The uplink
signal-to-interference ratio (SINR) of $i$-th user at $k$-th BS
is computed based on large scale fading coefficient as}

{
\[
\mathrm{SINR}_{ik}^{\mathrm{u}}\approx\frac{\beta_{ikk}}{\sum_{j\neq k}\beta_{ijk}},
\]
which is assumed to be known, e.g., based on the distances, or from
a radio map. The algorithm from \cite{zhusmart2015} considers a target
cell to be optimized, in the presence of already allocated pilot sequences
in surrounding cells. Starting from a random allocation in the target
cell, the algorithm identifies the user with minimum uplink SINR in
the target cell and then tries to improve its SINR by switching pilots
with another user in the target cell. This process is repeated until
convergence. As this procedure is not guaranteed to converge when
multiple cells are to be optimized simultaneously, we propose a modification,
given in Algorithm \ref{alg:Smart-pilot-algorithm}. }

\begin{algorithm}
\begin{enumerate}
\item {Initialize by randomly assigning users to pilots
such that every user in each cell is assigned to one pilot.}
\item {Sort all the users according to their uplink SINR
.}
\item {Take the user with the worst overall uplink SINR
(say user $i$ in cell $k$)}
\begin{enumerate}
\item {find another user in cell $k$ with best SINR to
possibly switch with, say $i$ }
\item {if after switching}
\begin{enumerate}
\item {the smallest SINR is not increased, then try to find
another user in cell $k$ and go to step (b). If there are no other
users available in cell $k$, STOP }
\item {the smallest SINR is increased, do the switch and
go to step 2}
\end{enumerate}
\end{enumerate}
\end{enumerate}
\caption{\label{alg:Smart-pilot-algorithm}Smart pilot assignment.}

\end{algorithm}

\subsection{{Heuristic Algorithm\label{subsec:Heuristic-Algorithm}}}

{In this subsection, a heuristic algorithm is proposed
to decrease the computational complexity of the proposed joint optimization
schemes. The proposed heuristic algorithm exploits not only distance
information but also AoA information. So, the algorithm assign users
to pilots based on the cost (\ref{eq:Uij}). Recall $U_{ij}$ is the
cost associated if $\ensuremath{i}$-th user and $j$-th user are
assigned to a pilot. The heuristic algorithm is similar in structure
to Algorithm \ref{alg:Smart-pilot-algorithm}. The main difference
is instead of uplink SINR, the cost is used for user assignment. The
proposed heuristic algorithm is detailed in Algorithm \ref{alg:Heuristic-algorithm}.
To improve the performance of the heuristic algorithm, once could
use the initial user assignment obtained from Algorithm \ref{alg:Smart-pilot-algorithm}.}

\begin{algorithm}
\begin{enumerate}
\item {Initialize by randomly assigning users to pilots
such that every user in each cell is assigned to one pilot.}
\item {Sort all the users according to their cost functions.
Note that the cost function captures both distance and AoA information.}
\item {Take the user with the highest overall cost (say
user $i$ in cell $k$)}
\begin{enumerate}
\item {find another user in cell $k$ with lowest cost to
possibly switch with, say $i$ }
\item {if after switching}
\begin{enumerate}
\item {the highest cost is not decreased, then try to find
another user in cell $k$ and go to step (b). If there are no other
users available in cell $k$, STOP }
\item {the highest cost is reduced, do the switch and go
to step 2}
\end{enumerate}
\end{enumerate}
\end{enumerate}
\caption{\label{alg:Heuristic-algorithm}Heuristic algorithm}

\end{algorithm}

\section{Numerical Results\label{sec:Numerical-Results}}

In this section, we present numerical results to evaluate the proposed
schemes described in Section \ref{sec:Coordinated-Pilot-Assignment}.

\subsection{Simulation Scenario}

\begin{table}
\centering

\caption{\label{tab:Simulation-Parameters}Simulation Parameters}

\begin{tabular}{|c|c|}
\hline
Parameter & Value\tabularnewline
\hline
\hline
{$\eta$} & {2.5}\tabularnewline
\hline
{$\lambda$} & {0.1 m}\tabularnewline
\hline
{$\sigma^{2}$} & {0.001}\tabularnewline
\hline
{$D$} & {$\lambda/2$}\tabularnewline
\hline
\end{tabular}\hspace{0.5 mm}%
\begin{tabular}{|c|c|}
\hline
Parameter & Value\tabularnewline
\hline
\hline
{$R$} & {1000 m}\tabularnewline
\hline
{$\gamma_{\mathrm{SNR}}$} & {20 dB}\tabularnewline
\hline
{$B$} & {50}\tabularnewline
\hline
{$P_{\mathrm{BS}}$} & {1W}\tabularnewline
\hline
\end{tabular}
\end{table}
We consider hexagonal shaped cells and simulations are performed with
a multi-cell system scenario. The simulation parameters used to obtain
the numerical results are given in Table \ref{tab:Simulation-Parameters}.
For a cell radius $R$, we set $\alpha$ to
\begin{equation}
\alpha[\textrm{dB}]=\gamma_{\mathrm{SNR}}+10\,\eta\,\log_{10}(R)+10\,\log_{10}(\sigma^{2}),
\end{equation}
where $\gamma_{\mathrm{SNR}}$ is the cell-edge SNR in dB and $\sigma^{2}$
is the receiver noise power variance.We keep these parameters fixed
for the simulations unless otherwise specified. {To
generate the AoA distributions, we consider a model where propagation
is dominated by a ring of scatterers around the user \cite{Guiseppe,Shiu:00,Abdi:02,Oestges:03,Oestges:04,Zhang:07}.
As an approximation, we consider a disk of radius $r_{ijk}$ comprising
many scatterers around the user $i$ in cell $j$. This radius can
be different for each user depending on the environment.  In that
case, $p(\theta_{ijk})$ corresponds to a distribution with support
$[\theta_{ijk}^{\mathrm{min}},\theta_{ijk}^{\mathrm{max}}]$, for
some fixed $\theta_{ijk}^{\mathrm{min}},\theta_{ijk}^{\mathrm{max}}\in[0,2\pi]$,
$\theta_{ijk}^{\mathrm{min}}<\theta_{ijk}^{\mathrm{max}}$ . We can
calculate $\theta_{ijk}^{\mathrm{min}}=\theta_{ijk}^{\mu}-\theta_{ijk}^{\delta}$,
$\theta_{ijk}^{\mathrm{max}}=\theta_{ijk}^{\mu}+\theta_{ijk}^{\delta}$
, where
\begin{align}
\theta_{ijk}^{\mu} & =\arctan\biggl(\frac{[\mathbf{x}_{ij}]_{2}-[\mathbf{x}_{k}]_{2}}{[\mathbf{x}_{ij}]_{1}-[\mathbf{x}_{k}]_{1}}\biggr),\label{eq:theta_mu}\\
\theta_{ijk}^{\delta} & =\arcsin\biggl(\frac{{r_{ijk}}}{\Vert\mathbf{x}_{ij}-\mathbf{x}_{k}\Vert_{2}}\biggr).\label{eq:theta_delta}
\end{align}
 As mentioned previously, we assume that a map exists in the base
station, connecting the user's position to the support of the AoA
distribution as well as the average received power (i.e., $\beta_{ijk}$
in the form of a radio map). From those maps, it is then possible
to compute $\mathbf{R}_{ijk}$, needed to  compute $\hat{\mathbf{h}}_{ikk}$
in (\ref{eq:MMSEinterference}). }\textcolor{red}{}
\begin{rem}
{We consider a non-LOS scenario with $B$ scattering
paths. However, there might be cases where the scatterers are strong,
causing large angular spreads, or cases where the location of users
is not accurately known (i.e., the location estimates have uncertainty).
These aspects can be incorporated into the system by allowing a larger
radius $r_{ijk}$, which translates to a larger range of angles\@.
More complex AoA distributions can also be used by means of multiple
scattering rings.}
\end{rem}

\subsection{{Performance Metrics}}

{Three different performance metrics are considered
to evaluate the proposed schemes with existing approaches.}
\begin{itemize}
\item {\emph{Normalized mean squared error of channel estimation}}{{}
(NMSE) of uplink channel estimate of user $i$ at the $k$-th BS in
$k$-th cell is denoted by $\eta_{ik}$}\textcolor{red}{}{{}
and is written as
\begin{equation}
\mathcal{\eta}_{ik}=\mathbb{E}\left\{ \frac{\Vert\hat{\mathbf{h}}_{ikk}-\mathbf{h}_{ikk}\Vert_{\mathrm{F}}^{2}}{\Vert\mathbf{h}_{ikk}\Vert_{\mathrm{F}}^{2}}\right\} ,\label{eq:err_metric}
\end{equation}
where $\mathbf{h}_{ikk}$ and $\hat{\mathbf{h}}_{ikk}$ are the desired
and estimated channels of user $i$ at the $k$-th BS in $k$-th cell
and the expectation is over many channel realizations. }
\item {\emph{Cumulative distribution function of channel
estimation errors:}}{{} Let us define the NMSE errors
in dB as $e_{ikk}=10\,\log_{10}(\frac{\Vert\hat{\mathbf{h}}_{ikk}-\mathbf{h}_{ikk}\Vert_{\mathrm{F}}^{2}}{\Vert\mathbf{h}_{ikk}\Vert_{\mathrm{F}}^{2}}$),
$e_{ikk}^{\textrm{no-int}}=10\,\log_{10}(\frac{\Vert\hat{{\bf h}}_{ikk}^{\textrm{no-int}}-\mathbf{h}_{ikk}\Vert_{\mathrm{F}}^{2}}{\Vert\mathbf{h}_{ikk}\Vert_{\mathrm{F}}^{2}})$,
of the estimated channels with and without interfering users respectively,
$\hat{{\bf h}}_{ikk}^{\textrm{no-int}}$ is the corresponding interference-free
channel estimate. We further denote $\epsilon_{ikk}=e_{ikk}-e_{ikk}^{\textrm{no-int}}$.
The channel estimation errors from all cells for each pilot $p$ is
collected in a set $\mathcal{E}_{p}=\{\epsilon_{ikk}\}_{k=1}^{L}$.
The cumulative distribution function (CDF) $F(\mathcal{E})$ is calculated
on $\mathcal{E}_{p}$ for every pilot based on several channel realizations.}
\begin{figure}
\includegraphics[width=1\columnwidth]{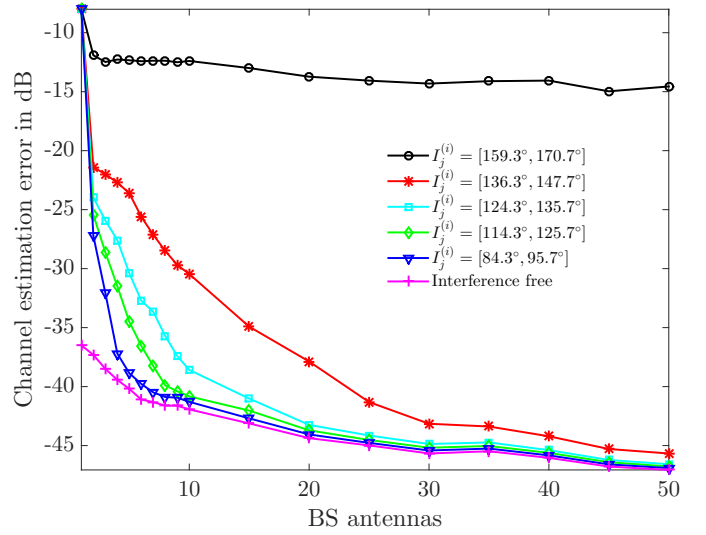}\caption{\label{fig:Impact_diff_mean_AoA}{Comparison of channel
estimation error $\eta$ versus BS antennas for different AoA $I_{j}^{(i)}$
of single interfering user. The target user is placed in $d=500$
m and the interfering user is placed in $d=1000$ m from the target
BS. The results are averaged over 1000 Monte-Carlo channel realizations. }}
\end{figure}
\item {\emph{Downlink per-cell sum rate:}}{{}
Assuming maximum ratio transmission,
{and the availability of perfect CSI at the user terminal, an upper bound on}
the downlink SINR of $i$-th user at $j$-th BS can be computed as
\begin{align*}
 & \textrm{SINR}_{ij}=\\
 & \frac{\vert\mathbf{h}_{ij}^{\mathrm{T}}\mathbf{w}_{ij}\vert^{2}}{\sum_{k\neq i}\vert\mathbf{h}_{ij}^{\mathrm{T}}\mathbf{w}_{kj}\vert^{2}+\sum_{l\neq j}\sum_{k=1}^{K}\vert\mathbf{h}_{il}^{\mathrm{T}}\mathbf{w}_{kl}\vert^{2}+\sigma_{n}^{2}\frac{MK}{P_{\mathrm{BS}}}}
\end{align*}
where $\mathbf{w}_{ij}=\frac{\hat{\mathbf{h}}_{ij}^{*}}{\Vert\hat{\mathbf{h}}_{ij}\Vert},\forall i,j$.
The factor in the numerator corresponds to received power, and the
first and second terms in the denominator capture the intra and inter
cell interference. The achievable downlink rate for the $i$-th user
in $j$-th BS is then can be calculated as $\Omega_{ij}=\log_{2}(1+\textrm{SINR}_{ij})$,
and the sum rate of all users in a cell is obtained $\Omega_{j}=\sum_{i=1}^{K}\Omega_{ij}$.
Finally the per-cell downlink sum rate is given by $\Omega=(1/L)\sum_{j=1}^{L}\Omega_{j}$.}
\begin{figure*}
\begin{centering}
\subfloat[]{\begin{centering}
\includegraphics[width=0.45\textwidth]{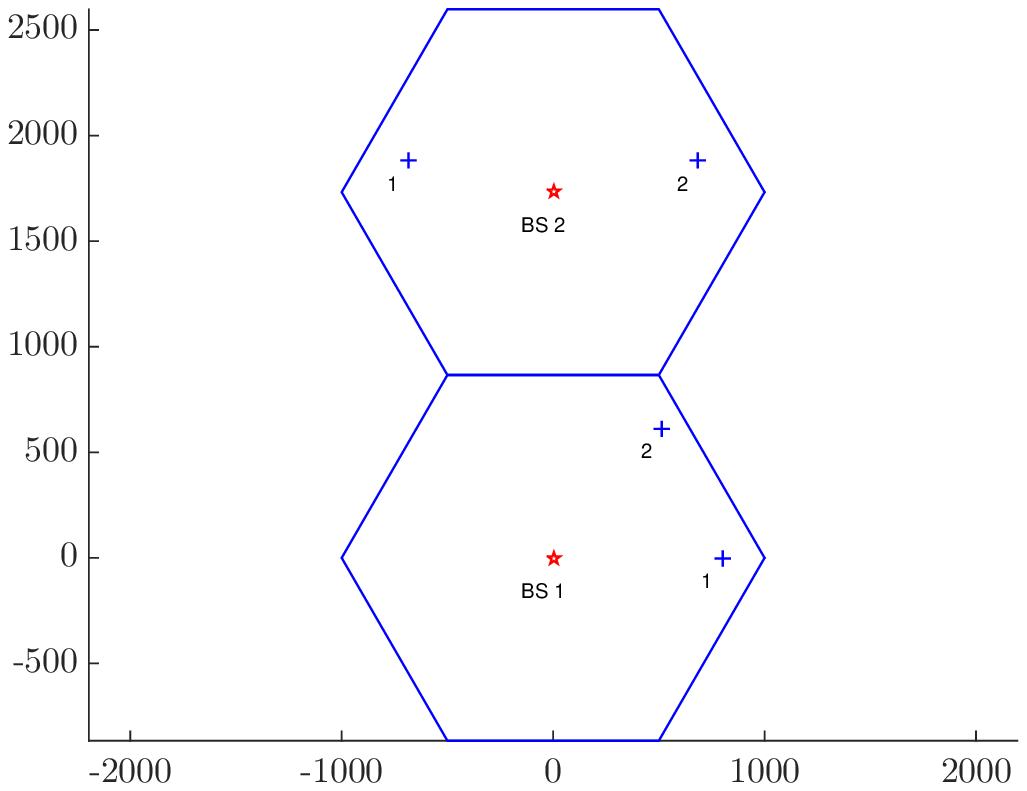}
\par\end{centering}
}\quad\subfloat[]{\begin{centering}
\includegraphics[width=0.45\textwidth]{GreedyVsJoint_2Cell}
\par\end{centering}
}
\par\end{centering}
\caption{\label{fig:Greedy_vs_JointOpt}Performance comparison of a greedy
vs joint optimization for a two cell scenario. Inset (a) two cell
scenario with two users each, users are marked with a plus sign and
BS with pentagons, and (b) channel estimation error $\eta$ as a function
of the number of BS for the users in cell 1 with greedy and joint
optimization schemes.}
\end{figure*}
\end{itemize}

\subsection{Results and Discussion}

\subsubsection{Impact of Different AoA Supports of an Interfering User}

\textcolor{red}{ }In Fig.~\ref{fig:Impact_diff_mean_AoA}, we show
the impact of different ranges of AoA support $I_{j}^{(i)}$ of a
single interfering user w.r.t. to $\textrm{DAR}_{i}$ of the target
user. Note that all the angles are after the axis transformation performed
w.r.t. to the target user. We can calculate $\textrm{DAR}_{i}$, and
for $M=10$, it is obtained as $[37.6^{\circ},142.4^{\circ}].$ We
varied $I_{j}^{(i)}$ such that some are within $\textrm{DAR}_{i}$
and some are outside this range. When $I_{j}^{(i)}$ lies within $\textrm{DAR}_{i}$
then the interference is low, and the channel estimation converges
fast to the interference-free scenario;\textbf{ }with $M=10$ BS antennas
the channel estimation performance is similar to that of the interference-free
scenario. On the other hand, when $I_{j}^{(i)}$ is outside the $\textrm{DAR}_{i}$,
it can be observed that more BS antennas are needed to converge to
the interference-free scenario. For example, when $I_{j}^{(i)}=[136.3^{\circ}147.7^{\circ}]$,
then more than 50 BS antennas are required.

\subsubsection{Two Cell Scenario: Single Cell Optimization}

We now look into the pilot allocation in a single cell, given known
allocations in other cells. In particular, we consider a two-cell
scenario with two users in each cell, $r_{ij}$ is set to 50 m for
each user, and $\tau=2$ (see Fig.~\ref{fig:Greedy_vs_JointOpt}
(a)), where the users in cell 2 have already been allocated pilots
and we aim to reuse these pilots for the users in cell 1. We compare
the proposed joint optimization scheme (\ref{multi_user_single_cell})
with greedy sequential user assignment \cite{muppirisetty2015Location,Gesbert}.
 For this scenario, both the users from cell 2 fall within the desired
angular region of user 1 in cell 1. However, for user 2 of cell 1,
only user 1 of cell 2 is permissible. In the greedy sequential scheme,
user 1 of cell 1 is assigned the same pilot as user 1 of cell 2, since
it provided the largest angular separation among the users in cell
2. Then user 2 of cell 1 has no choice but to be assigned the same
pilot as user 2 in cell 2. In contrast, the joint optimization considers
compatibility of \emph{both} users of cell 1. Therefore, it matches
user 1 of cell 1 with user 2 of cell 2 and user 2 of cell 1 with user
1 of cell 2. The channel estimation error $\mathcal{\eta}$ performance
for both schemes for the users in target cell (i.e., cell 1) is shown
in Fig.~\ref{fig:Greedy_vs_JointOpt} (b). As expected, the performance
of pilot 2 with greedy assignment suffers as user 2 of cell 1 is not
compatible with user 2 of cell 2.

\subsubsection{Two Cell Scenario: Multi-Cell Optimization}

\begin{figure}
\includegraphics[width=1\columnwidth]{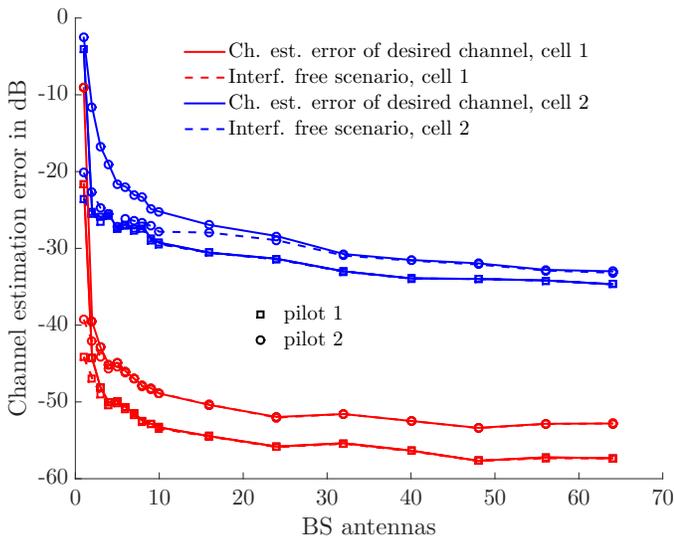}\caption{\label{fig:mutual_interf_2_cell}Channel estimation performance of
the target users in each cell for a two-cell scenario with joint optimization
scheme. The different colors show users in different cells, while
the markers show different pilots, reused in both cells. }
\end{figure}

We now extend the discussion for the scenario above to multi-user
multi-cell optimization (\ref{multi_user_multi_cell_qos}). Now we
can optimize the performance of both users in both cells. Therefore,
the joint optimization during user assignment considers not only reducing
the interference seen by the target user but also how the target user
is contributing interference to users in the neighboring cells. So,
users in different cells are assigned the same pilot if the AoA support
of a user in one cell lies within the desired angular region of another
user in another cell and vice-versa. Consequently, the channel estimation
performance of the target users at their respective BS is improved.
This is seen in Fig.~\ref{fig:mutual_interf_2_cell}, where the performance
of each user in each cell approaches the interference-free condition.

When we increase the number of users per cell to 5 and set $\tau=5$,
{$M=64$} and compare the {joint assignment from Section
\ref{subsec:Multi-User-Multi-Cell-Optimizati} with the smart pilot
algorithm \cite{zhusmart2015} described in Section \ref{subsec:Smart-Pilot-Algorithm},
and the proposed heuristic from Section \ref{subsec:Heuristic-Algorithm},
we obtain the results in Fig.~\ref{fig:CDF_2cell}. We have fixed
the users' locations and the ring sizes $r_{ij}$ (varying between
50 m and 100 m among users), and vary the channel and noise realizations
to obtain a CDF of the channel estimation errors. This implies that
the pilot assignments for each algorithms are fixed. {It can be observed that all the medthods
have similar performance when $M=64$ BS antennas. However, it was observed for when $M=20$
the proposed heuristic performs similar to that of joint optimization
scheme, while the smart pilot scheme leads to worse performance for
one of the pilots.}
\begin{figure}
{\includegraphics[width=1\columnwidth]{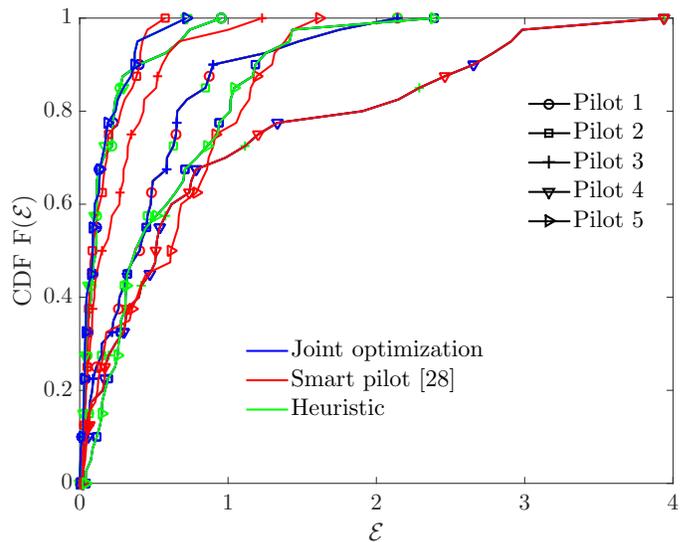}\caption{\label{fig:CDF_2cell}CDF of channel estimation error for 2 cell scenario
with 5 users per cell. For this scenario, $r_{ij}$ is chosen randomly
between {[}50,100{]} m for each user. The pilot length is $\tau=5$.
The BS antenna size is {$M=64$}. The CDF plot is compared for the each
pilot and for the joint optimization, smart pilot, and proposed heuristic
schemes.{{} }1000 Monte-Carlo channel realizations are
generated to obtain the results.{{} }}
}
\end{figure}

\textcolor{olive}{}

\subsubsection{Seven Cell Scenario: Multi-Cell Optimization}

In this section, we will evaluate the CDF of the{{}
channel estimation errors and the sum-rate for a seven-cell scenario,
with with 5 users per cell and a pilot length of $\tau=5$. Fig.~\ref{fig:CDF_5cell}
shows that according to the joint optimization method, there are two
bad pilots (i.e., a combination of users for which interference is
unavoidably high). For $\mathcal{E}<10$ dB, the heuristic outperforms
the smart pilot allocation, which has two bad pilots, compared to
one for the heuristic. When the number of cells is increased, it becomes
even more difficult to reduce the pilot contamination for all the
desired users in each cell per each pilot. The joint optimization
and heuristic approaches offer relatively better performance to smart
pilot scheme. }
\begin{figure}
{\includegraphics[width=1\columnwidth]{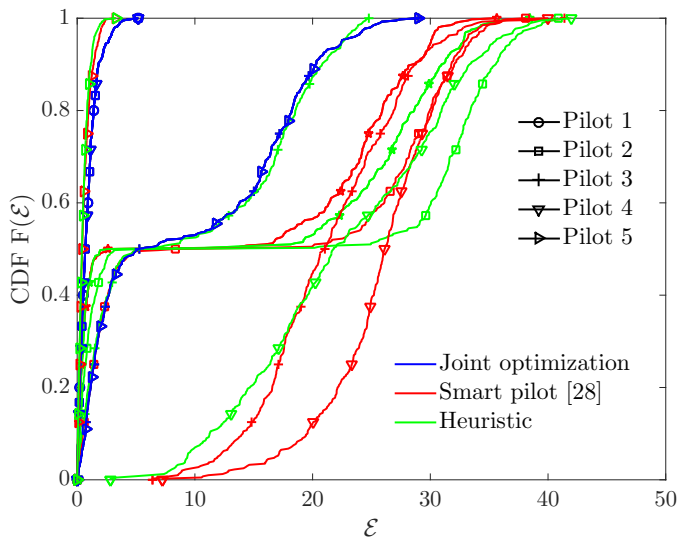}}

{\caption{\label{fig:CDF_5cell}CDF of channel estimation error for 7 cell scenario
with 5 users per cell. For this scenario, $r_{ij}$ is chosen randomly
between {[}50,100{]} m for the each user. The pilot length is $\tau=5$.
The BS antenna size is {$M=64$}. The CDF plot is compared for the each
pilot and for the joint optimization, smart pilot, and proposed heuristic
schemes. {1000 Monte-Carlo channel realizations are
generated to obtain the results.}}
}

\end{figure}

{In Fig. \ref{fig:per_cell_DL_increase_M}, the per-cell
downlink sum rate ($\Omega$) with increase in number of BS antennas
is depicted for 7-cell scenario with 5 users per cell. Furthermore,
we also consider a random user assignment scheme, where in users are
randomly assigned to pilots in each cell. { For the random scheme, the
average over the best sum rate over 1000 random pilot assignments is considered. Also, we depict the sum rate offered in the
case of interference free scenario.} We can clearly observe that
as the number of antennas is increased the per-cell downlink sum rate
is also increased. The joint optimization scheme offers better sum
rate compared to the other schemes. This is due to the fact that it
can better handle to avoid pilot contamination by choosing optimal
user allocation. For {$M=64$} BS antennas the joint optimization scheme
gives around 11\% more rate in comparison to random user assignment
scheme. {Furthermore, as $M$ increases the gap decreases between the sum rate offered by joint
optimization scheme and the interference free scenario.}}

{The offered sum rate by various schemes with increasing
number of users per cell is shown in Fig. \ref{fig:per_cell_DL_increase_users}.
The downlink sum rate increases with increase in number of users for
all the schemes, with a widening gap between the joint optimization
and the two competing methods. }

{}
\begin{figure}
{\includegraphics[width=1\columnwidth]{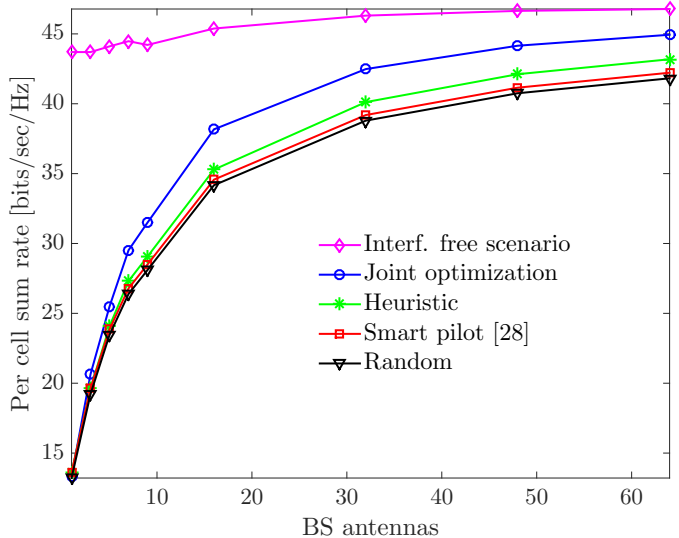}\caption{\label{fig:per_cell_DL_increase_M}The per-cell down link sum rate
as function of BS antennas.{{} 7-cell scenario} with
5 users per cell is considered. For this scenario, $r_{ij}$ is chosen
randomly between {[}50,100{]} m for the each user. The pilot length
is $\tau=5$. {The results are averaged over 1000
Monte-Carlo channel realizations.}}
}
\end{figure}
{}
\begin{figure}
{\includegraphics[width=1\columnwidth]{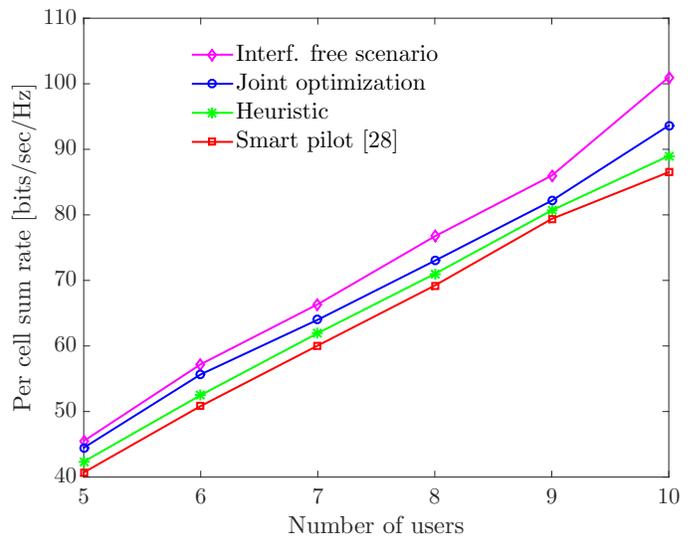}\caption{\label{fig:per_cell_DL_increase_users}The per-cell down link sum
rate as function of increasing users per cell for 7 cell scenario.
For this scenario, $r_{ij}$ is chosen randomly between {[}50,100{]}
m for the each user. The pilot length is varied from $\tau=5$ to
$\tau=10$. The BS antenna size is {$M=64$}. {The results
are averaged over 1000 Monte-Carlo channel realizations.}}
}
\end{figure}

\section{Conclusions and Future Directions\label{sec:Conclusions}}

In this paper, we characterized the effect of interference for MIMO
BSs with a large, but finite, number of antennas, harnessing user
location information. Building on this characterization, we formulated
several pilot assignment problems as integer quadratic constraint
optimization problem. These problems are solved centrally, provided
that all the information about all users is shared among all BSs.
 {To reduce the computational complexity of these
joint optimization problems, we further proposed heuristic algorithm
which assigns users to pilots based on both distance and angle of
arrival information of the users.} {We show proposed
pilot assignment strategies offer improved channel estimation performance
as well as enhanced downlink sum rate even when the number of antennas
is finite.} {However, low-complexity greedy methods
suffer from a severe performance penalty compared to joint optimization.
}

 Part of our ongoing research focuses on developing distributed implementations
{based on local information}.{{} Distributed
implementations for pilot contamination avoidance under a game theoretic
framework have been proposed using coalition games \cite{Mochaourab:2016}
and non-cooperative games \cite{ahmadi2016game}. However, none of
the aforementioned approaches exploits the location information that
BSs can obtain and considered in this work.  }

\appendices{}

\section{Channel estimation under finite MIMO\label{sec:Channel-estimation-finiteMIMO}}

{The channel estimate $\hat{\mathbf{h}}_{i}$ of the
desired user can be written using (\ref{eq:received_signal}) and
(\ref{eq:MMSEinterference}), with
\[
\mathbf{Q}_{i}=\mathbf{R}_{i}\Bigl(\sigma^{2}\mathbf{I}_{M}+\tau\sum_{j=1}^{L}\mathbf{R}_{j}\Bigr)^{-1}
\]
 as
\begin{align*}
\hat{\mathbf{h}}_{i} & =\mathbf{Q}_{i}\bar{\mathbf{S}}^{\mathrm{H}}(\bar{\mathbf{S}}\sum_{j=1}^{L}\mathbf{h}_{j}+\mathbf{n})\\
 & =\mathbf{Q}_{i}(\tau\mathbf{h}_{i}+\bar{\mathbf{S}}^{\mathrm{H}}\mathbf{n})+\tau\mathbf{Q}_{i}\sum_{j\neq i}\mathbf{h}_{j},
\end{align*}
{where $L=\vert\mathcal{C}\vert$ and the last terms constitutes the interference}. We approximate $\mathbf{R}_{j}$
by a low-rank version, i.e., $\mathbf{R}_{j}\approx\mathbf{U}_{j}\bm{\Sigma}_{j}\mathbf{U}_{j}^{\mathrm{H}}$
, in which $\bm{\Sigma}_{j}$ is an $m_{j}\times m_{j}$ matrix. In
general, for finite $M$, $\mathbf{R}_{j}$ is full rank, but for
a sufficiently large number of antennas, $m_{j}<M$, provided the
AoA support of $p(\theta_{j})$ is finite \cite{Gesbert}, so low-rank
approximation of $\mathbf{R}_{j}$ exists. Then, $\sum_{j=1}^{L}\mathbf{R}_{j}$
has a rank $m\le\sum_{j=1}^{L}m_{j}$ approximation:
\begin{align*}
\sum_{j=1}^{L}\mathbf{R}_{j} & \approx\sum_{j=1}^{L}\mathbf{U}_{j}\bm{\Sigma}_{j}\mathbf{U}_{j}^{\mathrm{H}}=\mathbf{U}\bm{\Sigma}\mathbf{U}^{\mathrm{H}}\\
 & =\mathbf{U}_{i}\tilde{\bm{\Sigma}}_{i}\mathbf{U}_{i}^{\mathrm{H}}+\mathbf{U}_{\bar{i}}\bm{\Sigma}_{\bar{i}}\mathbf{U}_{\bar{i}}^{\mathrm{H}},
\end{align*}
in which $\mathrm{span}(\mathbf{U})=\mathrm{span}(\mathbf{U}_{i})\cup\mathrm{span}(\mathbf{U}_{\bar{i}})$
is decomposed into two orthogonal spaces. Note that $\tilde{\bm{\Sigma}}_{i}$
may be different from $\bm{\Sigma}_{i}$. }

{We can now express $\mathbf{I}_{M}=\mathbf{V}\mathbf{V}^{\mathrm{H}}+\mathbf{U}\mathbf{U}^{\mathrm{H}}$,
where $\mathbf{V}$ is a unitary matrix spanning the orthogonal complement
of $\mathbf{U}\mathbf{U}^{\mathrm{H}}$. Then
\[
\mathbf{Q}_{i}=\mathbf{U}_{i}\bm{\Sigma}_{i}(\sigma^{2}\mathbf{I}_{m_{i}}+\tau\tilde{\bm{\Sigma}_{i}})^{-1}\mathbf{U}_{i}^{\mathrm{H}}.
\]
The interference from user $j$ is thus determined by
\[
\frac{\Vert\mathbf{U}_{i}\bm{\Sigma}_{i}(\sigma^{2}\mathbf{I}_{m_{i}}+\tau\tilde{\bm{\Sigma}_{i}})^{-1}\mathbf{U}_{i}^{\mathrm{H}}\tau\mathbf{h}_{j}\Vert^{2}}{\Vert\mathbf{U}_{i}\bm{\Sigma}_{i}(\sigma^{2}\mathbf{I}_{m_{i}}+\tau\tilde{\bm{\Sigma}_{i}})^{-1}\mathbf{U}_{i}^{\mathrm{H}}\tau\mathbf{h}_{i}\Vert^{2}}.
\]
Since $\mathbf{h}_{j}\in\mathrm{span}(\mathbf{a}(\theta_{j}))$, responses
$\mathbf{a}(\theta_{j})$ and $\mathbf{a}(\theta'_{j})$ for which
$\Vert\mathbf{U}_{i}^{\mathrm{H}}\mathbf{a}(\theta_{j})\Vert>\Vert\mathbf{U}_{i}^{\mathrm{H}}\mathbf{a}(\theta'_{j})\Vert$
indicate that $\mathbf{a}(\theta_{j})$ causes more interference.
Since $\mathbf{a}^{\mathrm{H}}(\theta_{j})\mathbf{R}_{i}\mathbf{a}(\theta_{j})=\mathbf{a}^{\mathrm{H}}(\theta_{j})\mathbf{U}_{i}\bm{\Sigma}_{i}\mathbf{U}_{i}^{\mathrm{H}}\mathbf{a}(\theta_{j})$,
it can be used as a measure of interference. Hence, interfering users
$j$ for which the steering vectors are such that $\mathbf{a}^{\mathrm{H}}(\theta_{j})\mathbf{R}_{i}\mathbf{a}(\theta_{j})$
is small are preferred over users for which $\mathbf{a}^{\mathrm{H}}(\theta_{j})\mathbf{R}_{i}\mathbf{a}(\theta_{j})$
is large, to limit the impact of interference during channel estimation
for user $i$.}

\section{Selection of $\tilde{\psi_{i}}^{\min}$ and $\tilde{\psi_{i}}^{\max}$\label{sec:Selection-of-}}

 We denote $\{\phi_{r,\theta_{i}^{\delta}}^{*}\}_{r}$ and $\{\phi_{r,\pi-\theta_{i}^{\delta}}^{*}\}_{r}$
as the sets of zeros of functions $J(\theta_{i}^{\delta},\phi)$ and
$J(\pi-\theta_{i}^{\delta},\phi)$, respectively. The values of $\phi_{\theta_{i}^{\delta}}^{*}$
and $\phi_{\pi-\theta_{i}^{\delta}}^{*}$ are obtained by solving
the following expressions using (\ref{eq:cosf})
\begin{align}
\cos(\phi_{\theta_{i}^{\delta}}^{*}) & =\cos(\theta_{i}^{\delta})+\frac{z\lambda}{MD},\label{eq:cos_phi_zero}\\
\cos(\phi_{\pi-\theta_{i}^{\delta}}^{*}) & =-\cos(\theta_{i}^{\delta})+\frac{z\lambda}{MD}.\label{eq:cos_phi_pi}
\end{align}
 We further define $\tilde{\psi_{i}}^{\min}=\max(\underset{r}{\min}(\{\phi_{r,\theta_{i}^{\delta}}^{*}\}_{r}),\underset{r}{\min}(\{\phi_{r,\pi-\theta_{i}^{\delta}}^{*}\}_{r}))$
and $\tilde{\psi_{i}}^{\max}=\min(\underset{r}{\max}(\{\phi_{r,\theta_{i}^{\delta}}^{*}\}_{r}),\max(\{\phi_{r,\pi-\theta_{i}^{\delta}}^{*}\}_{r}))$.

\section*{Acknowledgment}

We would like to thank Haifan Yin and David Gesbert for fruitful discussions.
The authors would like to thank GUROBI for the free academic license
to use their numerical optimization software.

\bibliographystyle{IEEEtran}
\bibliography{references}

\end{document}